# Strong dipolar effects in a quantum ferrofluid

Thierry Lahaye[1], Tobias Koch[1], Bernd Fröhlich[1], Marco Fattori[1], Jonas Metz[1], Axel Griesmaier[1], Stefano Giovanazzi[1] & Tilman Pfau[1]

[1] 5. Physikalisches Institut, Universität Stuttgart, Pfaffenwaldring 57, D-70550 Stuttgart, Germany

**Symmetry-breaking interactions play a crucial role in many areas of physics, ranging from classical ferrofluids to superfluid $^3$He and d-wave superconductivity. For superfluid quantum gases, a variety of new physical phenomena arising from the symmetry-breaking interaction between electric or magnetic dipoles are expected[1]. Novel quantum phases in optical lattices, such as checkerboard or supersolid phases, are predicted for dipolar bosons[2,3]. Dipolar interactions can also enrich considerably the physics of quantum gases with internal degrees of freedom[4,5,6]. Arrays of dipolar particles could be used for efficient quantum information processing[7]. Here we report on the realization of a Chromium Bose-Einstein condensate (BEC) with strong dipolar interaction. By using a Feshbach resonance, we reduce the usual isotropic contact interaction, such that the anisotropic magnetic dipole-dipole interaction between $^{52}$Cr atoms becomes comparable in strength. This induces a change of the aspect ratio of the cloud, and, for strong dipolar interaction, the inversion of ellipticity during expansion – the usual "smoking gun" evidence for BEC – can even be suppressed. These effects are accounted for by taking into account the dipolar interaction in the superfluid hydrodynamic equations governing the dynamics of the gas, in the same way as classical ferrofluids can be described by including dipolar terms in the classical hydrodynamic equations. Our results are a first step in the exploration of the unique properties of quantum ferrofluids.**

A quantum ferrofluid is a superfluid quantum gas consisting of polarized dipoles, either electric or magnetic. The first option might be achieved for instance with polar molecules in their vibrational ground state, aligned by an electric field. Progress has been made recently in slowing and trapping of polar molecules[8], but the densities and temperatures achieved to date are far away from the quantum-degenerate regime. The use of Feshbach resonances to create polar molecules from two ultracold atomic species[9] is a promising, actively explored alternative[10]; however it is a challenging task to bring those heteronuclear molecules to their vibrational ground state[11]. Alternatively, atomic electric dipoles induced by dc fields[12] or by light[13] could be used. The other option, chosen here, relies on the magnetic dipole-dipole interaction (MDDI) between atoms with a large magnetic moment, such as Chromium, for which a BEC was achieved recently[14]. The relative strength of the MDDI to the contact interaction is conveniently expressed by the dimensionless ratio

$$\varepsilon_{dd} = \frac{\mu_0 \mu^2 m}{12\pi \hbar^2 a} , \qquad (1)$$

where $m$ is the atomic mass, $a$ the s-wave scattering length and $\mu$ the magnetic moment (numerical factors in $\varepsilon_{dd}$ are such that a homogenous BEC with $\varepsilon_{dd} > 1$ is unstable against dipolar collapse). Chromium has a large dipole moment $\mu = 6\,\mu_B$ and a background scattering length in the fully polarized case $a \approx 100\,a_0$ ($\mu_B$ is the Bohr magneton; $a_0$ the Bohr radius), yielding[15] $\varepsilon_{dd} \approx 0.16$. Although this value is typically 36 times larger than in standard alkali quantum gases, the MDDI is still a small perturbation compared to the contact interaction. A perturbative mechanical effect of the MDDI has been demonstrated by analyzing the expansion of a Chromium BEC from an anisotropic trap for various orientations of the dipoles[16].

The existence of Feshbach resonances[17] allows to increase $\varepsilon_{dd}$ and go beyond the perturbative limit. Indeed, close to a resonance, the scattering length varies with the applied magnetic field $B$ as

$$a = a_{bg}\left(1 - \frac{\Delta}{B - B_0}\right) \qquad (2)$$

where $a_{bg}$ is the background scattering length, and $\Delta$ the resonance width. For $B$ approaching $B_0 + \Delta$, the scattering length tends to zero, thus enhancing $\varepsilon_{dd}$. This gives the possibility to reach a MDDI-dominated quantum gas.

We report here on the observation of strong dipolar effects in a Chromium BEC in the vicinity of the broadest Feshbach resonance at $B_0 \approx 589$ G. We measure the dispersive behaviour of the scattering length, and observe that the change in $a$ is accompanied by enhanced inelastic losses. Close to the zero-crossing of $a$, we observe a large modification of the aspect ratio of the cloud when $\varepsilon_{dd}$ increases, which is direct evidence for strong dipolar effects. We finally show that the usual inversion of ellipticity of the condensate during expansion is inhibited for large enough MDDI.

We modified our experimental setup, which has been described in detail elsewhere[14], in order to be able to produce Cr condensates in high field, close to $B_0$ (see Methods summary). Once the condensate is obtained, the magnetic field is ramped close to its final value $B$ in 10 ms, and held there for 2 ms to let it settle down. The trap is then switched off, and the condensate expands freely for 5 ms before being imaged by absorption of resonant light in high field. Figure 1a shows a series of images taken when approaching the resonance from above, and clearly displays a reduction of the cloud size, as well as a change in its aspect ratio.

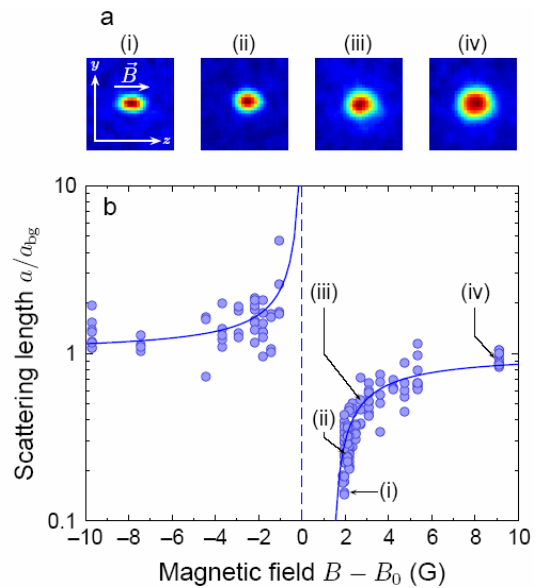

**Figure 1. Tuning the Chromium scattering length. a:** Absorption images (field of view 260 μm by 260 μm) of the condensate after 5 ms of expansion, for different fields $B$ above resonance ($B - B_0$ is 2, 2.2, 2.7 and 9 G from (i) to (iv)). Reducing $a$ slows down the mean-field driven expansion. The change in aspect ratio for small $a$ is a direct signature of strong MDDI. **b:** Variation of $a$ across the resonance, inferred from the mean-field energy released during expansion. The line is a fit to Eqn. (2), yielding $\Delta = 1.4 \pm 0.1$ G.

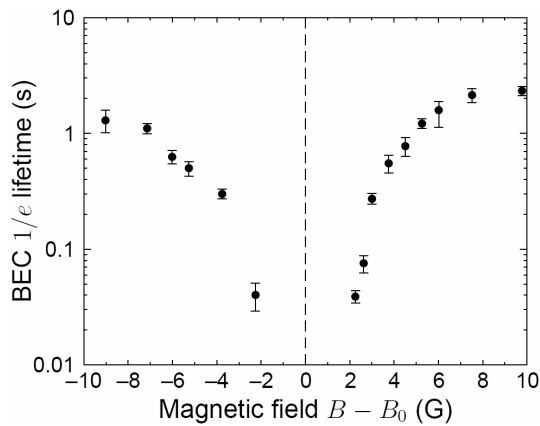

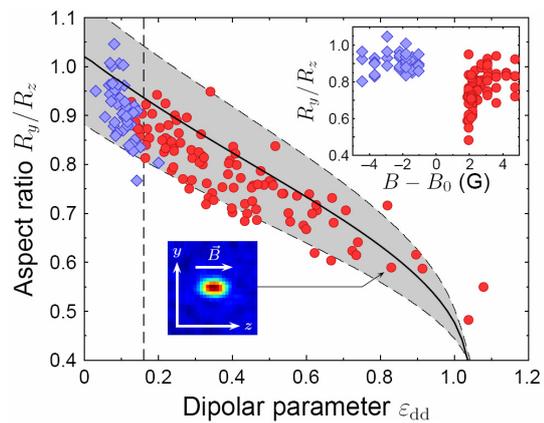

**Figure 2. Inelastic losses close to the resonance.** The $1/e$ lifetime of the condensate is plotted as a function of the magnetic field $B$. Error bars correspond to one standard deviation in the lifetime determination using a fit of the atom number remaining in the BEC after a variable holding time, by an exponential decay law. The losses are small enough to allow for the observation of enhanced dipolar interactions (see Methods).

**Figure 3. Increasing the dipolar parameter.** Condensate aspect ratio after 5 ms of expansion, versus the measured $\varepsilon_{dd}$. Diamonds (circles) correspond to data taken below (above) resonance. The solid line is the prediction of hydrodynamic theory including MDDI, without adjustable parameter (the grey-shaded area corresponds to the uncertainties in the trap frequencies). The dashed line indicates the off-resonant $\varepsilon_{dd}$ value. Inset: subset of the same data points plotted versus the magnetic field. The condensate elongates appreciably along $B$ only just above resonance, when $a$ approaches zero. The sample absorption image gives an example of condensate with strong MDDI.

From the measured optical density profiles, we extract the BEC atom number $N$, as well as the Thomas-Fermi radii $R_z$ (along the magnetization direction) and $R_y$ (along the vertical axis). Without MDDI, one would easily obtain the scattering length $a$ from these measurements, since the Thomas-Fermi radii after time of flight would scale as $(Na)^{1/5}$. In our case, we take into account the MDDI using the hydrodynamic formulation of the Gross-Pitaevskii equation, including both contact and dipole-dipole interactions[18] (see Methods summary for the assumptions underlying our analysis). Figure 1b shows the measured variation of $a(B)$ across the resonance, showing a characteristic dispersive shape[19]. A fit according to Eqn. (2) yields $\Delta = 1.4 \pm 0.1$ G, in good agreement with the prediction $\Delta = 1.7$ G of multi-channel calculations[17]. The position $B_0 \approx 589$ G of the resonance coincides with the one obtained by observing inelastic losses in a thermal cloud[16]. We can tune $a$ by more than one order of magnitude, with a reduction by a factor of five above the resonance.

Close to the resonance, we observe on both sides enhanced inelastic processes resulting in a decay of the condensate. We studied the BEC atom number as a function of the time spent at the final magnetic field $B$, and fitted the corresponding curves by an exponential decay law (this functional form being chosen for simplicity). The $1/e$ BEC lifetime obtained in this way is shown in figure 2 as a function of $B$ (the initial peak atomic density is $3 \times 10^{14}$ cm$^{-3}$). Enhanced inelastic losses close to a Feshbach resonance have been observed with other species, e.g. sodium[20]. Here the losses are small enough to allow for the observation of the enhanced relative strength of the MDDI on the equilibrium shape of the condensate.

Figure 3 shows the aspect ratio $R_y/R_z$ of the cloud as a function of the value $\varepsilon_{dd}$ obtained from the measured $a$ and constitutes the main result of this Letter. The aspect ratio decreases when $\varepsilon_{dd}$ increases: the cloud becomes more elongated along the direction of magnetization $z$, as can be seen unambiguously in Figure 1a. This is a clear signature of the MDDI, as for pure contact interaction, the aspect ratio is independent of the scattering length (provided the Thomas-Fermi approximation is valid). The solid line in Figure 3 shows the aspect ratio after time of flight calculated using the hydrodynamic theory including MDDI[18], without any adjustable parameter. The agreement between our data and the theoretical prediction is excellent, given the dispersion of data points and the uncertainty in the theoretical prediction arising from the trap frequencies measurements. The highest value of $\varepsilon_{dd}$ we could reach reliably is about 0.8, corresponding to a five-fold reduction of the scattering length. For our trap geometry, the condensate is expected to become unstable with respect to dipolar collapse[21,22] for values of $\varepsilon_{dd}$ slightly above one (the exact value depending on the trap anisotropy, but also on the atomic density).

As an application of the tunability of the dipolar parameter, we study the expansion of the condensate for two orthogonal orientations of the dipoles with respect to the trap, as was done in Ref. 16, but now as a function of $\varepsilon_{dd}$. In practice, for the large fields required to approach the Feshbach resonance, we cannot change the magnetic field orientation, which is always along $z$. We therefore use two different trap configurations, with interchanged $y$ and $z$ frequencies, and identical frequencies along $x$: trap 1 has frequencies $(\omega_x,\omega_y,\omega_z) / 2\pi \approx (660,370,540)$ Hz, while trap 2 has $(\omega_x,\omega_y,\omega_z) / 2\pi \approx (660,540,370)$ Hz. We then measure the aspect ratio of the cloud (defined as $A_1 = R_z/R_y$ for trap 1, and $A_2 = R_y/R_z$ for trap 2) as a function of the time of flight, for different values of $B$ (and hence of $\varepsilon_{dd}$). This protocol is equivalent to a mere rotation of the magnetization direction with respect to the trap axes.

Figure 4 presents the results. In order to check that the two trap configurations only differ by an exchange of the $y$ and $z$ frequencies, we first perform the expansion experiment without switching on the large magnetic field along $z$, but only a small (11.5 G) field along $x$ (the line of sight of the imaging). The magnetization is therefore perpendicular to the observation plane, and changing the trap configuration does not affect the aspect ratios, as the difference between the two situations is simply a rotation of the trap *around the magnetization axis*. Figure 4a shows the equality $A_1 = A_2$, confirming the equivalence of the two configurations. We then study the expansion with the large magnetic field $B$ along $z$. In this case, the MDDI induces a change in the aspect ratios, and $A_1 \neq A_2$. Far from resonance, $\varepsilon_{dd} \approx 0.16$ and we recover the perturbative dipolar effect already observed in Ref. 16 (see Figure 4b). However, for values of $B$ approaching $B_0 + \Delta$, $\varepsilon_{dd}$ increases and, correspondingly, the difference between $A_1$ and $A_2$ becomes very large (Fig. 4c and 4d, where $\varepsilon_{dd}$ is $0.5 \pm 0.1$ and $0.75 \pm 0.1$, respectively). The lines correspond to the prediction of the hydrodynamic theory without adjustable parameter, and show again a very good agreement with the data. The effect of the dipolar interaction is way beyond the perturbative regime, and induces very strong deviations from what one expects for contact interaction. In particular, for $\varepsilon_{dd} \approx 0.75$, the aspect ratio $A_2$ always remains smaller than unity during the time of flight: the strong anisotropic dipolar interaction inhibits the inversion of ellipticity, the "smoking gun" evidence for BECs with contact interaction.

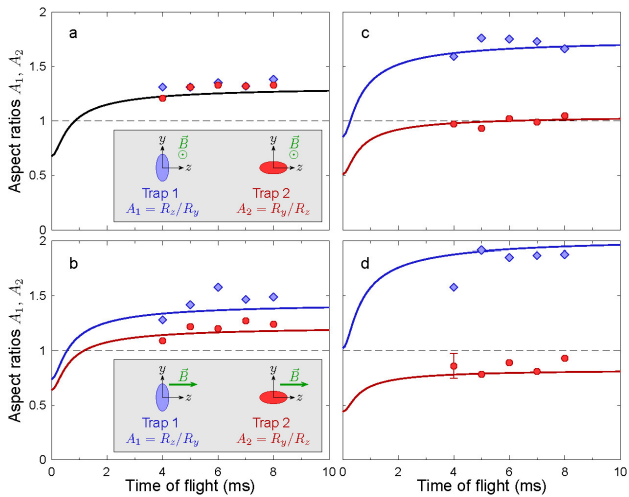

**Figure 4. MDDI-dominated BEC expansion.** Aspect ratio of the condensate versus time of flight, for two traps with interchanged $y$ and $z$ frequencies (see text). The solid lines are theoretical predictions without adjustable parameter. The error bar (panel d) gives the typical dispersion over several runs. a: Dipoles perpendicular to the observation plane (see inset sketching the in-trap BEC shape); both configurations yield the same aspect ratio. b, c, d: Dipoles along $z$ (see inset in b); $\varepsilon_{dd}$ takes the values 0.16, 0.5, and 0.75. The MDDI induces larger and larger effects, even inhibiting (d) the inversion of ellipticity.

In conclusion, the use of a Feshbach resonance to reduce the s-wave scattering length of Chromium allowed us to realize a BEC with strong dipolar interaction, and study the hydrodynamics of this novel quantum ferrofluid. This work opens up many avenues towards the study of dipolar quantum gases beyond the perturbative regime. Structured density profiles are predicted for dipolar condensates in anisotropic traps[23], including biconcave density distributions in pancake-shaped traps[24]. A clear direction for future work is thus to use a one-dimensional optical lattice creating a stack of pancake-shaped traps. A condensate with dipoles perpendicular to the trap plane is then stable with respect to dipolar collapse, which should allow to enter the regime $\varepsilon_{dd} \gg 1$. In particular, the investigation of the unusual, roton-like excitation spectrum predicted in this system[25] is a fascinating perspective. The creation of $Cr_2$ molecules by ramping over the Feshbach resonance is another appealing experiment. In a two-dimensional trap, the repulsive interaction between the molecules, due to their large magnetic moment, might stabilize them against inelastic losses. Another possible extension of this work is the study of degenerate fermions with strong dipolar interactions, which may display new types of pairing mechanisms[1]. Finally, the behaviour of strongly correlated dipolar quantum gases in three-dimensional optical lattices is a fascinating open field with many connections to fundamental questions in condensed-matter physics, such as the study of supersolid phases, whose experimental observation in helium is still debated[26].

**Methods summary**

We modified our experimental sequence[14] to produce Chromium condensates in high field. For this, we switch on quickly (in less than 5 ms) a large field (~ 600 G) during forced evaporation in the dipole trap. The low atomic density at this stage of evaporation allows for small losses. The current in the coils used to produce the field is actively stabilized; care is taken to ensure a high homogeneity of the field. Evaporation is then resumed until an almost pure condensate of $3 \times 10^4$ atoms is obtained. The trap is then adjusted to obtain frequencies $(\omega_x, \omega_y, \omega_z) / 2\pi \approx (840, 600, 580)$ Hz (measured by exciting the centre of mass motion of the cloud, with an accuracy of 5%).

In our data analysis to extract the scattering length $a$ (Figure 1), we assumed that no external forces act on the atoms during the time of flight, that the condensate stays in equilibrium during the magnetic field ramp, and finally that the hydrodynamic (Thomas-Fermi) approximation is valid even for small $a$. These assumptions are largely fulfilled for all our parameters.

**Full methods** accompany this paper.

**Acknowledgements** We thank Luis Santos for stimulating discussions and Jürgen Stuhler for his contributions in the initial phases of the experiment. We acknowledge financial support by the German Science Foundation (SFB/TRR 21 and SPP 1116) and the EU (Marie-Curie fellowship to T.L.).

**Author information** The authors declare no competing financial interests. Correspondence and requests for materials should be addressed to T.L. (t.lahaye@physik.uni-stuttgart.de) or T.P. (t.pfau@physik.uni-stuttgart.de).

**Methods**

**Production of Cr condensates in a high magnetic field.** We load Chromium atoms in the $|^7S_3, m_S = 3\rangle$ state into a Ioffe magnetic trap for ten seconds, and subsequently cool them by rf-induced evaporation down to 20 μK. We capture $10^6$ of these atoms in an optical trap consisting of a horizontal laser beam at 1076 nm with a power of 16 W and a $1/e^2$ radius of 30 μm, and optically pump them in the high-field seeking state $|m_S = -3\rangle$. A 11.5 G magnetic field prevents losses due to dipolar relaxation. At this stage, a vertical beam (power 9 W, $1/e^2$ radius 50 μm) is ramped up over 5 s, creating a `dimple' in the trap. We then ramp the horizontal beam power down to 30% of its initial value in 5 s, and switch on rapidly (less than 5 ms) a magnetic field $B_{\text{evap}}$ along z. The field is provided by the offset and pinch coils of the Ioffe trap, in which we run currents around 400 and 15 A, respectively. This combination ensures that the field is as homogenous as possible. The remaining inhomogeneities correspond to trapping (anti-trapping) frequencies below 5 Hz (resp. 7 Hz) radially (resp. longitudinally). The offset coils current is stabilized at the $3 \times 10^{-5}$ level (rms), giving in principle a control of the field better than 100 mG. $B_{\text{evap}}$ is 600 G (resp. 575) for the data taken above (resp. below) resonance. The fast magnetic ramp and the low atomic density (about $10^{13}$ cm$^{-3}$) at this stage of the evaporation are required to cross the resonances below $B_{\text{evap}}$ without appreciable losses. Forced evaporation is resumed in high field until an almost pure condensate of $3 \times 10^4$ atoms is obtained. The power of the beams are finally adjusted adiabatically to reach the desired trap frequencies.

**Data analysis for the measurement of $a$.** The assumptions in our analysis are: (i) no external forces act on the atoms during the expansion; (ii) the condensate stays in equilibrium during the magnetic field ramp; and (iii) the hydrodynamic (Thomas-Fermi) approximation is valid. We therefore checked that (i) the effect of the inhomogeneities of the magnetic field during expansion is completely negligible for our parameters; (ii) the magnetic field ramp is slow enough to fulfil the adiabaticity criterion $\dot{a}/a \ll \omega_{\min}$, where $\omega_{\min}$ is the smallest trap frequency. We checked that no collective oscillations were excited in the condensate when varying $a$. For this, we varied the time spent at the final magnetic field $B$ before expansion, and observed no change in the condensate shape, even for the data taken close to $B_0 + \Delta$. Note that the occurrence of losses makes the adiabaticity criterion more difficult to fulfil, as one must stay far enough from resonance in order to satisfy $\dot{N}/N \ll \omega_{\min}$. This criterion is largely fulfilled for our data. Finally, we checked (iii) that the hydrodynamic description of the condensate is valid: the Thomas-Fermi parameter, defined as $Na/a_{\text{ho}}$, where $a_{\text{ho}} = \sqrt{\hbar/(m\bar{\omega})}$ is the trap harmonic oscillator length and $\bar{\omega} = (\omega_x \omega_y \omega_z)^{1/3}$, is always higher than $\sim 60$ (a value achieved for the lowest values of $a$ and $N$). The Thomas-Fermi condition $Na/a_{\text{ho}} \gg 1$ is therefore fulfilled.